\documentclass[twocolumn,preprintnumbers,amsmath,amssymb,showpacs,superscriptaddress,prb,aps,10pt,floatfix]{revtex4-1}

\usepackage{graphicx}
\usepackage{bm}
\usepackage{dcolumn}
\usepackage{amssymb}
\usepackage{amsmath}
\usepackage{amsbsy}
\usepackage{amsfonts}
\usepackage{epsfig}
\usepackage{graphicx}
\usepackage{stackrel}
\usepackage{paralist}
\usepackage[rgb]{xcolor}
\usepackage{todonotes}
\usepackage{pdfcomment}
\usepackage[percent]{overpic}
\usepackage{tikz}
\usetikzlibrary{intersections} 
\usetikzlibrary{positioning,calc}
\usepackage[caption=false]{subfig}

\tikzset{
    right angle quadrant/.code={
        \pgfmathsetmacro\quadranta{{1,1,-1,-1}[#1-1]}     
        \pgfmathsetmacro\quadrantb{{1,-1,-1,1}[#1-1]}},
    right angle quadrant=1, 
    right angle length/.code={\def\rightanglelength{#1}},   
    right angle length= 1 ex, 
    right angle symbol/.style n args={3}{
        insert path={
            let \p0 = ($(#1)!(#3)!(#2)$) in     
                let \p1 =  ($(\p0)!\quadranta*\rightanglelength!(#3)$),
                \p2 = ($(\p0)!\quadranta*\rightanglelength!(#2)$) in 
                let \p3 = ($(\p1)+(\p2)-(\p0)$) in  
            (\p1) -- (\p3) -- (\p2)
        }
    }
}

\newcommand{\I}{\mathrm{i}}

\newcommand{\refEq}[1]{Eq.~(\ref{#1})}
\newcommand{\refFig}[1]{Fig.~\ref{#1}}
\newcommand{\refSctn}[1]{Section~\ref{#1}}

\newcommand{\citeRef}[1]{Ref.~\onlinecite{#1}}
\newcommand{\refApp}[1]{Appendix~\ref{#1}}

\definecolor{NewColor}{rgb}{1,0,0}

\newcommand{\CO}[1]{\textcolor{red}{}}

\newlength{\noLengthl}
\newcommand{\nolength}[1]{\ensuremath{#1\settowidth{\noLengthl}{$#1$}\hspace*{-\noLengthl}}}
\newcommand{\intS}[2]{\ensuremath{\int_{\nolength{#1}}^{\nolength{#2}}}}

\newcommand{\sumS}[2]{\ensuremath{\settowidth{\noLengthl}{$#1$}\hspace*{-#2 \noLengthl} \sum_{#1} \hspace*{- #2 \noLengthl}}}
\renewcommand{\Im}{\mathrm{Im}}
\renewcommand{\Re}{\mathrm{Re}}

\renewcommand{\vec}[1]{{\boldsymbol{#1}}}

\begin{document}
\title{
\texorpdfstring{Second quantization of Leinaas-Myrheim anyons in one dimension \\ and their relation to the Lieb-Liniger model}{Second quantization of Leinaas-Myrheim anyons in one dimension and their relation to the Lieb-Liniger model}
}
\author{Thore Posske}

\affiliation{I. Institut f{\"u}r Theoretische Physik, Universit{\"a}t Hamburg, Jungiusstra{\ss}e 9, 20355 Hamburg, Germany}

\author{Bj{\"o}rn Trauzettel}
\affiliation{Institute  for  Theoretical  Physics  and  Astrophysics, University  of  W{\"u}rzburg, 97074  W{\"u}rzburg, Germany}
\author{Michael Thorwart}
\affiliation{I. Institut f{\"u}r Theoretische Physik, Universit{\"a}t Hamburg, Jungiusstra{\ss}e 9, 20355 Hamburg, Germany}

\begin{abstract}
In one spatial dimension, anyons in the original description of Leinaas and Myrheim are formally equivalent to locally interacting bosons described by the Lieb-Liniger model.  
This admits an interesting reinterpretation of interacting bosons in the context of anyons.
We elaborate on this parallel, particularly including the many-body bound states from the attractive Lieb-Liniger model.
In the anyonic context these bound states are created purely by quantum-statistical attraction and coined quantum-statistical condensate, which is more robust than the Bose-Einstein condensate.
We introduce the second quantization formalism for the present anyons and construct the generalized Jordan-Wigner transformation that connects them to the bosons of the Lieb-Liniger model.
\end{abstract}
\pacs{ 
05.30.Pr 	
03.65.Ge, 	
71.10.Pm 	
}
\keywords{anyons, Lieb-Liniger model, interacting bosons, second quantization, many-body, exact, anyon condensate, statistical interaction, differential equations, Robin boundary conditions}
\maketitle 

\section{Introduction}
In modern nanoscopic systems, electronic excitations are effectively confined to a lower-dimensional world. 
An unexpected consequence of such a reduced spatial dimension is the 
occurrence of particles that neither obey Fermi nor Bose statistics: anyons \cite{%
	LeinaasMyrheim1977TheoryOfIdenticalParticles,%
	Kretzschmar1965MustQuantalWaveFunctionsBeSingleValued,%
	LaidlawDeWitt1971FeynmanFunctionalIntegralsForSystemsOfIndistinguishableParticles,%
	Dowker1972QuantumMechanicsAndFieldTheoryOnMultiplyConnectedAndOnHomogeneousSpaces,%
	GoldinMenikoffSharp1980ParticleStatisticsFromInducedRepresentationsOfALocalCurrentGroup,%
	Wilczek1982QuantumMechanicsOfFractionalSpinParticles%
}.
In two dimensions, anyons have been theoretically extensively studied \cite{BiedenharnLiebSimonWilczek1990TheAncestryOfTheAnyon,khare2005FractionalStatisticsAndQuantumTheory,Kitaev2006AnyonsInAnExactlySolvableModelAndBeyond} and indicated to exist in several experimental systems %
\cite{%
	MooreRead1991NonabelionsInTheFractionalQuantumHallEffect,%
	Laughlin1983AnomalousQuantumHallEffectAnIncompressibleQuantumFluidWithFractionallyChargedExcitations,%
	Ivanov2001NonAbelianStatisticsOfHalfQuantumVortices,%
	CaminoZhouWeiGoldman2005RealizationOfALaughlinQuasiparticleInterferometerObservationOfFractionalStatistics,%
	WilletPfeifferWest2010AlternationAndInterchangeOfEo4andEo2PeriodInterferenceOscillationsConsistentWithFillingFactor5o2NonAbelianQuasiparticles,%
ZhongXuWangSonGuoLiuXuXuaLuHanPanWang2016EmulatingAnyonicFractionalStatisticalBehaviorInASuperconductingQuantumCircuit
}.
The spatial exchange of two-dimensional anyons and the accompanied, fixed unitary transformation of the anyonic wave function 
could, amongst others, pave the way to topological quantum computing \cite{NayakSternFreedmanDasSarma2008NonAbelianAnyonsAndTopologicalQuantumComputation}. 
This exchangeability is also apparent in a single spatial dimension by considering ringlike systems or T-structures
\cite{ParkRecher2015DetectingTheExchangePhaseOfMajoranaBoundStatesInACorbinoGeometryTopologicalJosephsonJunction,AliceaOregRefaelOppen2011NonAbelianStatisticsAndTopologicslQuantumInformationProcessingIn1DWireNetworks}.
Sparked by this idea, the interest in lower-than-two-dimensional anyons has recently risen, especially in conjunction with the possible detection of Majorana bound states in quantum wires %
\cite{%
	TangEggertPelster2015GroundStatePropertiesOfAnyonsInA1DLattice,%
	KeilmannLanzmichMcCullockRoncaglia2011StatisticallyInducedPhaseTransitionsAndAnyonsIn1DOpticalLattices,%
	NadjPergeDrozdovLiChenJeonSeoMacDonaldBernevigYazdani2014ObservationOfMajoranaFermionsInFerromagneticAtomicChainsOnASuperconductor,%
	MourikZuoFrolovPlissardBakkersKouwenhoven2012SignaturesOfMajoranaFermionsInHybridSuperconductorSemiconductorNanowireDevice,%
	Kitaev2001UnpairedMajoranaFermionsInQuantumWires%
}.
Those are expected to be non-abelian anyons with potential application to topological quantum computing \cite{AliceaOregRefaelOppen2011NonAbelianStatisticsAndTopologicslQuantumInformationProcessingIn1DWireNetworks, HalperinBertrandOregSternRefaelAliceaOppen2012AdiabaticManipulationsOfMajoranaFermionsInA3DnetworkOfWires}.

Within this work, we employ the concept of anyons
introduced in the seminal 
work by Leinaas and Myrheim \cite{LeinaasMyrheim1977TheoryOfIdenticalParticles}.
Their approach advantageously bases only on one fundamental idea: to set up the proper classical theory of indiscernible particles and then quantize it.
In two dimensions, this results in ``standard'' Chern-Simons anyons, imaginable as bosons with an attached flux acquiring an Aharonov-Bohm phase when physically exchanged \cite{Wilczek1982QuantumMechanicsOfFractionalSpinParticles,Kitaev2006AnyonsInAnExactlySolvableModelAndBeyond}. This renders Leinaas' and Myrheim's approach one of the standard references in the field.
Less applied is their theory to one spatial dimension. Here their approach competes with manifold theories 
\cite{Gentile1940OsservazioniSopraLeStatisticheIntermedie,Fendley2012ParafermionicEdgeZeroModesInZNInvariantSpinChains,AliceaFendley2016TopologicalPhasesWithParafermionsTheoryAndBlueprints,DaiXie2009IntermediateStatisticsSpinWaves,HutterWoottonLoss2015ParafermionsInAKagomeLatticeOfQuibitsForTopologicalQuantumComputation,GoldinMenikoffSharp1980ParticleStatisticsFromInducedRepresentationsOfALocalCurrentGroup,	GoldinMenikoffSharp1981RepresentationsOfALocalCurrentAlgebraInSonsimplyConnectedSpaceAndTheAharonovBohmEffect,LeinaasMyrheim1988IntermediateStatisticsForVorticesInSuperfluidFilms,Hansson1992DimensionalReductionInAnyonSystems,Haldane1991FractionalStatisticsInArbitraryDimensionsAGeneralizationOfThePauliPrinciple,Ha1995FractionalStatisticsInOneDimensionViewFromAnExactlySolvableModel,Pasquier1994ALectureOnTheCalogeroSutherlandModels,FisherGlazman1996TransportInA1DLuttingerLiquid,BatchelorGuanOelkers2006OneDimensionalInteractingAnyonGasLowEnergyPropertiesAndHaldaneExclusionStatistics, Kundu1999ExactSolutionOfDoubleDeltaFunctionBoseGasThroughAnInteractingAnyonGas,BrinkHanssonVasiliev1992ExplicitSolutionToTheNBodyCalogeroProblem,HaoZhangChen2009GroundStatePropertiesOfHardCoreAnyonsIn1DOpticalLattices,TangEggertPelster2015GroundStatePropertiesOfAnyonsInA1DLattice}; see 
\refApp{appDifferentFormalismsForAnyons}
for a brief summary.
In particular, the Leinaas-Myrheim anyons have to be contrasted to the quasi particle excitations of the Calogero-Sutherland model, the Haldane-Shastry chain, and the fractional excitations in Tomonaga-Luttinger liquids \cite{Pasquier1994ALectureOnTheCalogeroSutherlandModels,BrinkHanssonVasiliev1992ExplicitSolutionToTheNBodyCalogeroProblem,FisherGlazman1996TransportInA1DLuttingerLiquid,BatchelorGuanOelkers2006OneDimensionalInteractingAnyonGasLowEnergyPropertiesAndHaldaneExclusionStatistics, Kundu1999ExactSolutionOfDoubleDeltaFunctionBoseGasThroughAnInteractingAnyonGas,TangEggertPelster2015GroundStatePropertiesOfAnyonsInA1DLattice} that are as well called anyons \cite{BatchelorGuanOelkers2006OneDimensionalInteractingAnyonGasLowEnergyPropertiesAndHaldaneExclusionStatistics}.
The defining property of these kinds of anyons is that the wave function acquires the fixed phase $\kappa$ when the coordinates of two anyons get permuted, in complete analogy to the statistical angle in two spatial dimensions. 
While this behavior is deemed as natural, Leinaas and Myrheim emphasize that within their framework the axiom of impenetrable anyons is essential, which a priori prohibits the physical exchange of particles in one dimension. 
Still, these theories are directly incorporable into the framework of Leinaas and Myrheim by a proper continuation procedure and, as given in \citeRef{BatchelorGuanOelkers2006OneDimensionalInteractingAnyonGasLowEnergyPropertiesAndHaldaneExclusionStatistics}, by replacing the statistical parameter $\eta \rightarrow \eta / \cos(\kappa/2)$.
To strengthen the reasonability of Leinaas' and Myrheim's approach in sight of the other ones, we want to stress its fundamentality. First, it does not require an underlying theory of additional, constituting particles.  Only the indiscernibility of the considered particles, the validity of canonical quantization for flat spaces, and the hermiticity of the Hamiltonian is assumed.
Secondly, Leinaas-Myrheim anyons appear naturally when two-dimensional Chern-Simons anyons are confined to one dimension by a potential. 
In the process of the dimensional crossover, the complete statistical angle gets gradually absorbed and encoded into the scattering behavior of the anyons \cite{Hansson1992DimensionalReductionInAnyonSystems}, which eliminates the need for an additional statistical phase in one spatial dimension.
As a concrete physical situation, we might imagine a fractional quantum Hall insulator \cite{Laughlin1983AnomalousQuantumHallEffectAnIncompressibleQuantumFluidWithFractionallyChargedExcitations} where anyonic bulk excitations are confined to one dimension by an electric potential.

As a twist of history, Leinaas' and Myrheim's theory of one-dimensional anyons (1977) turn out to be formally equivalent to the Lieb-Liniger model, describing $\delta$-function interacting bosons in one dimension (1963) \cite{LiebLiniger1963}.
In fact, Lieb and Liniger derived the solutions of their model by first rewriting it employing boundary conditions, not knowing that these equations would several years later be reused by Leinaas and Myrheim to describe one-dimensional anyons.
While the Lieb-Liniger model has substantially advanced since, the results have, to the best of our knowledge, not been carried over to the regime of Leinaas-Myrheim anyons.
In this work, we close this gap.
This includes the calculation of observables \cite{HaimBergOppenOreg2015CurrentCorrelationsInAMajoranaBeamSplitter,RosenowLevkivskyiHalperin2016CurrentCorrelationsFromAMesoscopicAnyonCollider,TangEggertPelster2015GroundStatePropertiesOfAnyonsInA1DLattice,KeilmannLanzmichMcCullockRoncaglia2011StatisticallyInducedPhaseTransitionsAndAnyonsIn1DOpticalLattices,StraeterSrivastavaEckardt2016FloquetRealizationAndSignaturesOf1DAnyonsInAnOpticalLattice}
for confined anyons: the energy spectrum, momentum density, and finite size density oscillations. 
Our results are readily applicable to quasi-particle excitations in quasi one-dimensional systems, like interacting cold atom/ion chains and edge liquids of topological insulators, that potentially carry anyonic excitations %
\cite{%
	KeilmannLanzmichMcCullockRoncaglia2011StatisticallyInducedPhaseTransitionsAndAnyonsIn1DOpticalLattices,%
	TangEggertPelster2015GroundStatePropertiesOfAnyonsInA1DLattice,%
	HaoZhangChen2009GroundStatePropertiesOfHardCoreAnyonsIn1DOpticalLattices,%
	FuKane2008SuperconductingProximityEffectAndMajoranaFermionsAtTheSurfaceOfATopologicalInsulator,%
	LevinStern2009FractionalTopologicalInsulators%
}. 
On the other hand, we employ the interpretation of interacting bosons as anyons to bring original perspectives and approaches into the well established field of the Lieb-Liniger model.
For instance, the results for confined anyons also describe Lieb-Liniger bosons in a box, where the Dirichlet boundary conditions result in modified Bethe ansatz equations \cite{Gaudin1971BoundaryEnergyOfABoseGasIn1D, HaoZhangLiangChen2006GroundStatePropertiesOf1DUltracoldBoseGasesInAHardWallTrap}.
Additionally, we develop the second quantization formalism for Leinaas-Myrheim anyons, a step which is conceptually purely motivated by the anyonic interpretation.
While presenting the formalism, we take particular care to include complex momenta, which, as usual, describe spatially bound states.
In the exact many-body solutions, they build up the quantum-statistical condensate, a remarkably stable quantum phase. 
The existence of the bound states for a certain range of the statistical scattering parameter serves a little mystery in the anyonic interpretation, but immediately becomes clear as this range of the statistical parameter is analogue to the attractive regime of the Lieb-Liniger model.
Lieb and Liniger have first disregarded this regime as unphysical and unstable \cite{LiebLiniger1963}. More recent work has albeit revealed its soundness \cite{MugaSnider1998SolvableThreeBosonModelWithAttractiveDeltaFunctionInteraction,MontinaArecchi2005ManyBodyGroundStatePropertiesOfAnAttractiveBoseEinsteinCondensateInA1DRing,KanamotoSaitoUeda2003QuantumPhaseTransitionIna1DBoseEinsteinCondensateWithAttractiveInteractions, SakmannStreltsovAlonCederbaum2005ExactGroundStateOfFiniteBoseEinsteinCondensateOnARing, SykesDrummondDavis2007ExcitationSpectrumOfBosonsInAFinite1DCircularWaveguideViaTheBetheAnsatz}.
Recently, it has been pointed out that within the attractive regime additional gas-like phases may exist \cite{BatchelorBortzGuanOelkers2005EvidenceForTheSuperTonksGirardeauGas,AstrakharchikBoronatCasullerasGiogini2005BeyondTheTonksGirardeauGasStronglyCorrelatedRegimeInQuasi1DBoseGases}.

The structure of the paper is as follows. In \refSctn{sctnModel} we concisely review both relevant models, i.e., the Lieb-Liniger model and the model of Leinaas and Myrheim for one dimensional anyons and state their formal equivalence. In \refSctn{sctnConstructionOfWaveFunction}, we construct the anyonic wave functions, which are the basis of the second quantization formalism that we derive in \refSctn{sctnSecondQuantization}. We also provide the generalized Jordan-Wigner transformation from the Lieb-Liniger bosons to the Leinaas-Myrheim anyons. The Bethe ansatz equations for systems of finite size are discussed in \refSctn{sctnFiniteSizeSystems}, which are consequently, in \refSctn{sctnApplication}, applied to derive some properties of anyons in a box. The case of a negative statistical parameter is covered in \refSctn{sctnTheQuantumStatisticalCondensate}, where we introduce the quantum-statistical condensate and the interpretation of the clusters as individual anyons themselves. We conclude our work in \refSctn{sctnConclusions}.

\section{Model}
\label{sctnModel}
We start by reviewing the models of Lieb-Liniger and Leinaas and Myrheim and highlight their equivalence.
The Lieb-Liniger model describes a number of $n$ $\delta$-function interacting bosons in one dimension. In real space, the system is represented by its totally symmetric wave function $\Psi$, which maps from $\mathbb{R}^n$ to $\mathbb{C}$, and is governed by the Hamiltonian
\begin{align}
\label{eqnHamiltonian}
\mathcal{H}_{LL} = -\frac{\hbar^2}{2 m} \sum_{j=1}^{n} \partial^2_{x_j}
+ 2 c \sum_{i\neq j}  \delta\left(x_i - x_j \right). 
\end{align}
Here, $m$ denotes the mass of the particles and $c$ is the real interaction strength that has units of momentum. 
As familiar from elementary quantum mechanics, the $\delta$-functions can be directly implemented into the wave function by demanding boundary conditions, which, because of the symmetry of the wave function, turn out to be the so-called Robin boundary conditions
\begin{align}
\label{eqnImpermeabilityConstraint}
	\left( 
		\partial_{x_{j+1}}-\partial_{x_j}
	\right) 
	\Psi(\vec{x})
\mid_{x_{j} \to x_{j+1}}
&= 
	c \ \Psi(\vec{x}) 
\mid_{x_{j} \to x_{j+1}}, 
\end{align}
for each $j$ between $1$ and $n-1$,
and we restrict ourselves to the region $\mathcal{R} = \left\{ \vec{x} \mid x_1<x_2<\dots<x_n \right\}$ of the parameter space \cite{LiebLiniger1963}.
In exchange for the boundary conditions, the Hamiltonian on $\mathcal{R}$ becomes free, i.e., 
\begin{align}
\mathcal{H}_{LL}\mid_\mathcal{R} = -\frac{\hbar^2}{2m}\sum_{j=1}^n \partial_{x_j}^2.
\label{eqnFreeHamiltonian}
\end{align}

Let us now concisely recapitulate and slightly extend the quantum theory of Leinaas and Myrheim \cite{LeinaasMyrheim1977TheoryOfIdenticalParticles} for indiscernible particles.
First, consider $n$ classical particles in a region $M$ of real space.
The spatial configurations of a system of discernible particles would be described by tuples of positions $\vec{x} = \left(x_1, \dots, x_n\right)$, where $x_j$ lies in $M$.
Because the particles are indiscernible, however, using tuples is prodigal: for $n=2$, $(x_1,x_2)$ and $(x_2,x_1)$ label the same configuration.
Instead, we employ the sets $\left\{x_1, \dots, x_n\right\}$ of $n$ distinct positions. The family of all these sets is called configuration space $\mathcal{R}$ and inherits various properties by local equivalence to $M^n$.
Here, the notational correspondence of the configuration space to the parameter region of the Lieb-Liniger model is on purpose, since
for $n$ indiscernible particles on a line, the real space variables $x_1 < \dots <x_n$ parametrize $\mathcal{R}$.
To obtain the quantum mechanical theory,
space and momentum variables get promoted to the usual operators
acting on the wave functions $\Psi:\mathcal{R}\to \mathbb{C}$.
Additionally, the Hamiltonian $\mathcal{H}$ must be hermitian.
We consider, for compliance, the particles to also obey the free Hamiltonian of \refEq{eqnFreeHamiltonian},
however, electro-magnetic potentials and particle interactions can be added without changing the general formalism. 
Interestingly, hermiticity is exactly granted if $\Psi$ fulfills the Robin boundary conditions of \refEq{eqnImpermeabilityConstraint}. In this context, the interaction strength $c$ is called the statistical parameter $\eta \equiv c$.

We conclude that both theories use the same differential equation and boundary conditions, which constitutes a formal equivalence between them.
The twofold interpretation, however, provides simple explanations of seemingly complicated facts.
For instance, the boundary conditions of \refEq{eqnImpermeabilityConstraint} reduce to the Neumann and Dirichlet boundary conditions of bosons (at $c=0$) and fermions ($c = \pm \infty$) \footnote{To avoid divergences, {\protect$\eta = \pm \infty$} is defined as  {\protect$\frac{1}{\eta} = 0$}. Before taking the limit, {\protect$\eta$} has to be brought to the left hand side of {\refEq{eqnImpermeabilityConstraint}}. In subsequent equations, the limit of $\eta$ is taken before values for other variables are plugged in.
}. Hence, we immediately understand that hard core bosons, described by the Lieb-Liniger model with $c \to \infty$, behave like fermions, described by the model of Leinaas and Myrheim with $\eta \to \infty$. 
Also, if the bosons attract, i.e., $c<0$, they form clusters \cite{CalabreseCaux2007CorrelationFunctionsOfThe1DAttractiveBoseGas}.
This intuitively clear fact in the Lieb-Liniger model seem surprising for Leinaas-Myrheim anyons.
\footnote{
There also is a vivid explanation why anyons can bind that does not leave the anyonic point of view. To this end, note that a scattering event of two anyons with a negative statistical parameter results in a negative time shift of the corresponding matter wave \cite{Hansson1992DimensionalReductionInAnyonSystems}. 
Thus, anyons effectively scatter backwards in time, have the chance to scatter again, and form a bound state by infinite repetition.
}

There is indication that the coincidentally looking equivalence between $\delta$-interacting bosons and Leinaas-Myrheim anyons is in fact not coincidental.
To explain this, we refer to the connection between bosons and anyons in two spatial dimensions. There, anyons are equivalent to bosons that acquire an Aharonov-Bohm flux when circling around each other, hence, also in two dimensions, anyons are equivalent to particularly interacting bosons. The reason for this is that the configuration space for two-dimensional anyons has holes, the points where the position of a pair of particles would coincide. The holes themselves are irrelevant concerning scattering since particles can move around them by infinitesimally altering their path. However, the holes potentially induce a holonomy in the wave function, the Aharonov-Bohm phase. 
In one spatial dimension, the holes no longer induce a holonomy because particles cannot be exchanged. However, the holes themselves become relevant as, by normal propagation, particles are destined to scatter at some time. Then, the holes induce the boundary condition of \refEq{eqnImpermeabilityConstraint} and thereby again serve as the origin of the bosonic interaction that connects bosons and anyons.

In the remaining course of the paper, we present and interpret the solutions to the equivalent models from the more rarely employed anyonic point of view.

\section{Construction of the wave functions}
\label{sctnConstructionOfWaveFunction}
\begin{figure}
\newcommand{\scale}{0.4}

\newcommand{\scaleC}{0.2}
\def\deltaX{0.75}
\def\circleScale{0.1}
\def\deltaA{0.3}
\def\deltaHA{0.35}
\def\deltaB{0.2}
\def\deltaHB{0.25}
\def\base{-0.6}
\def\dist{1.5}

\begin{center}
\begin{tabular}{cccc}
&\multicolumn{1}{l}{\textbf{(a)}}	&\multicolumn{1}{l}{\textbf{(b)}}	&\multicolumn{1}{l}{\textbf{(c)}}	
\vspace*{-3 ex}
\\
\begin{tikzpicture}[scale=0.75]
\draw (3,\base) node {$\Re\left\{\vec{k}\right\}$};
\draw (3,\base-0.5) node {$\Im\left\{\vec{k}\right\}$};
\end{tikzpicture}
&
\hspace*{2ex}
\begin{tikzpicture}[scale=0.75]
				\begin{scope}[shift={(-4,0)}]
					\foreach \x in {1} {
					\draw [fill] (\x,0) circle (\circleScale);
					}
					\draw [blue] (1-\deltaA,0) -- (1-\deltaA,\deltaHA) --  node[anchor=south] {$\vec{\mu^1}$} (1+\deltaA,\deltaHA) -- (1+\deltaA,0);

					\draw (1,\base) node {$\left(K^1\right)$};
					\draw (1,\base-0.5) node {$\left(0\right)$};
				\end{scope}
				\end{tikzpicture}
&
\hspace*{1 ex}
\begin{tikzpicture}[scale=0.75]
				\begin{scope}[shift={(-4,0)}]
					\foreach \x in {1,2} {
					\draw [fill] (\x,0) circle (\circleScale);
					}
					\draw [blue] (1-\deltaA,0) -- (1-\deltaA,\deltaHA) --  node[anchor=south] {$\vec{\mu^2}$} (2+\deltaA,\deltaHA) -- (2+\deltaA,0);
					
					\draw (1.5,\base) node {$\left(K^2,K^2\right)$};
					\draw (1.5,\base-0.5) node {$-\frac{\eta}{2} \left(-1,1\right)$};
				\end{scope}
				\end{tikzpicture}
&
	\begin{tikzpicture}[scale=0.75]
				\begin{scope}[shift={(-5,0)}]
					\foreach \x in {1,2,3,4} {
					\draw [fill] (\x,0) circle (\circleScale);
					}
					\draw [blue] (1-\deltaA,0) -- (1-\deltaA,\deltaHA) --  node[anchor=south] {$\vec{\mu^5}$} (4+\deltaA,\deltaHA) -- (4+\deltaA,0);
					\draw (2.5,\base) node {$\left(K^5,K^5,K^5,K^5\right)$};
					\draw (2.5,\base-0.5) node {$-\frac{\eta}{2} \left(-3,-1,1,3\right)$};
				\end{scope}
				\end{tikzpicture}
\end{tabular}
	
\end{center}
\caption{%
\label{figGraphicalRepresentationOfBoundStates}
Examples of composite anyons described by 
irreducible clusters $\vec{\mu}$, i.e., strings of the Lieb-Liniger model. 
These are the fundamental building blocks of the anyonic wave functions and should be conceived as individual particles.
For clusters of more than one anyon, $\eta<0$ is implicit.
{(a)} single anyon; {(b)} two-anyon bound state; 
{(c)} maximally bound cluster of four anyons: the quantum-statistical condensate's ground state if generalized to $n$ particles.
}
\end{figure}
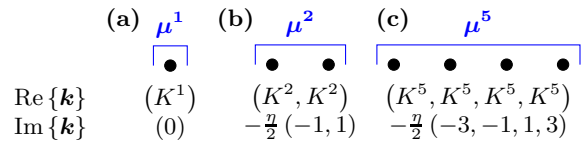

We next construct all wave functions that fulfill \refEq{eqnImpermeabilityConstraint}, combining the solutions of the attractive and the repulsive Lieb-Liniger model \cite{LiebLiniger1963, CalabreseCaux2007CorrelationFunctionsOfThe1DAttractiveBoseGas}, with the final aim to provide the second quantization formalism for the particles at hand.
To this end, we employ the ansatz \cite{LiebLiniger1963}  
$
\Psi\left(\vec{x}\right) = \int_{\vec{k} \in \mathbb{C}^n}{} d^n {k} \ \alpha\left(\vec{k}\right) e^{\I \vec{k}\vec{x}}. 
$
Complex momenta are explicitly included.
These are needed to describe the mentioned anyonic bound states that form for a negative statistical parameter.
In momentum space, the boundary conditions translate to 
\begin{align}
\label{eqnAlphaConstraint}
\alpha\left(\vec{k} \right) &= 
	e^{-\I \phi_\eta({k_{j+1}-k_j})} \alpha \left(\sigma_j \vec{k} \right) & \text{if $k_{j+1}-k_{j} \neq \I \eta$},
	\\
	\label{eqnCuttingCondition}
\alpha\left(\vec{k} \right) &= 		
	0  & \text{if $k_{j+1}-k_{j} = \I \eta$}.
\end{align}
Here, $\sigma_j$ denotes the elementary permutation which permutes the $j^{th}$ and $(j+1)^{th}$ element of a tuple and
\begin{align}
\label{eqnStatisticalPhase}
\phi_\eta({k_{j+1}-k_j}) = {2 \arctan\left[{\eta}/({k_{j+1}-k_{j})}\right]}
\end{align}
is the statistical phase.
By iteration, these conditions connect coefficients of relatively permuted momenta $\alpha(\vec{k}) = e^{\I \phi_\eta^P(\vec{k})} \alpha(P \vec{k})$.
Here, if $P= \sigma_{j_1}  \dots  \sigma_{j_r}$ 
is a general permutation written with an $r$ as small as possible, then 
\mbox{%
$
\phi^P_\eta\left(\vec{k}\right) 
	= \sum_{i=1}^r \phi_\eta \left[ 
					\left(
						\sigma_{j_1} \dots \sigma_{j_{i}} 
						\vec{k} 
					\right)_{j_i}
					-\left(
						\sigma_{j_1} \dots \sigma_{j_{i}} 
						\vec{k}
					\right)_{j_i+1}
				\right]
$}.
The basis functions are therefore of the form
\mbox{%
$
\Psi_{\vec{k}} \propto \sum_{P \in S_n } e^{\I \phi_\eta^P(\vec{k})} e^{\I \left(P  \vec{k}\right) \vec{x}}%
$}.
Divergent elements of this set are excluded by only permitting $\vec{k}$ being the concatenation 
\footnote{
The concatenation of tuples means to write their elements in a row to form a combined tuple, i.e., 
$
\vec{k} = \left(
					\vec{\mu^1},
					\dots, \vec{\mu^m}
		\right) = \left(\mu^1_1, \mu^1_2, \dots, \mu^m_{n_{{m}}-1},{\mu^m_{n_{m}}}\right)%
$,
if $\vec{\mu^1}, \dots, \vec{\mu^m}$ are tuples of lengths $n_1$, \dots, $n_m$ and $\vec{k}$ is their concatenation.
}
of tuples  
$\vec{\mu}$ 
of complex momenta where the difference $\mu_{j+1}-\mu_j$ between adjacent momenta is $-\I\eta$. 
These tuples are called strings in the context of the Lieb-Liniger model \cite{CalabreseCaux2007CorrelationFunctionsOfThe1DAttractiveBoseGas}.
Within the anyonic context, we call them clusters.
Examples of clusters are sketched in \refFig{figGraphicalRepresentationOfBoundStates}.
Physically, clusters with more than one element represent composite anyons whose constituents move colletively and are localized within a characteristic length scale of $1/\eta$ from each other. They should be conceived as individual particle themselves.
For positive $\eta$, irreducible clusters only consist of single particles, which describes free, unbound, anyons.
In order to uniquely label the basis functions, we introduce the cluster ordering $\mathcal{O}$. This is done in direct analogy to the ordering that needs to be introduced to label fermionic basis states in the standard many-body theory \cite{AltlandCondensedMatter}. 
To apply $\mathcal{O}$ to a tuple $\mathcal{D}$ of clusters, first merge clusters that are not disjoint to larger clusters by taking their union (and reorder to obtain the form of a cluster). Then, sort the resulting clusters 
by real part. Finally, concatenate the clusters in this order to obtain the tuple $\mathcal{O}\left(\mathcal{D}\right)$. 

In conclusion, the basis wave functions describe composite and free anyons in momentum space.
Given an ordered tuple $\mathcal{O}\left(\mathcal{D}\right)$ of irreducible clusters, the corresponding basis function obtains the form
\begin{align}
\label{eqnBasisFunction}
 \Psi_{{\bm{(}}k=\mathcal{O} 
	\left(
		\mathcal{D}\right)\bm{)}}\left(\vec{x}
	\right) = N_{\vec{k}} \sum_{P \in S_n}
 e^{
  \I
    \phi^P_\eta\left(
		      \vec{k}
    \right)
   }
  e^{    \I \left(
	      P\vec{k}
      \right) \vec{x}
 },
 \end{align}
where
$N_{\vec{k}}$ is the normalization \footnote{For real vectors $\vec{k}$, we have {\protect$N_{\vec{k}} = \sqrt{1/\left(2 \pi^n n!\right)}$}.} %
and
$e^{\I \phi^P_\eta\left(\vec{k}\right)}$ plays the role of a generalized Slater determinant. 

\section{Second quantization}
\label{sctnSecondQuantization}
An advantage of the anyonic interpretation is that a second quantization of the solution is reasonably motivated. In contrast, this endeavor seems to be discouraged in the bosonic picture of the Lieb-Liniger model. The formalism can facilitate the calculation of diverse properties, similar as the original second quantization of bosons and fermions does.
Given the basis wave functions of \refEq{eqnBasisFunction}, second quantization amounts to defining creation operators to construct all basis states from a vacuum state \footnote{Technically, we define the vacuum state as $\Psi_{\{\}} = 1$.
}.
For a cluster $\vec{\mu}$, we define its creation operator by
\begin{align}
\label{eqnSecondQuantizationCreationOperator}
a^\dagger_{\vec{\mu}}  \Psi_{\mathcal{O}\left(\mathcal{D}\right)} =&  
\sqrt{n_{\vec{\mu}} +1}\ 
e^{\I \Phi_\eta^{\vec{\mu}}\left(\mathcal{D}\right)}
\Psi_{\mathcal{O}\left(\left\{\vec{\mu}\right\} \cup  \mathcal{D}\right) }
\end{align}
and linear continuation to all states.
Here, $n_{\vec{\mu}}$ is the number of clusters $\vec{\mu}$ in $\mathcal{D}$. %
The phase
$
\Phi_\eta^{\vec{\mu}}\left(\vec{D}\right) =
\sum_{ 
	\vec{\tilde{\mu} <\vec{\mu}} 
}
 \varphi_\eta^{\vec{\tilde{\mu}},\vec{\mu}}%
$
is composed of the cluster-cluster exchange phases
$\varphi_\eta^{\vec{\tilde{\mu}},\vec{\mu}} = \sum_{i=1}^{{N({\vec{\mu}})}}
\sum_{i=j}^{{N({\tilde{\vec{\mu}}})}}
\phi_\eta(\tilde{\mu}_j - \mu_i)$. %
Here, $\vec{\tilde{\mu}}<\vec{{\mu}}$ if $\vec{\tilde{\mu}}$ is ordered to the left of $\vec{{\mu}}$ by cluster ordering and $N(\vec{\mu})$ denotes the number of anyons in $\vec{\mu}$.
Employing \refEq{eqnSecondQuantizationCreationOperator}, the algebra of the cluster creation operators is 
\begin{align}
\label{eqnQuantumAlgebraClusters}
a^\dagger_{\vec{\mu_1}} a^\dagger_{\vec{\mu}_2} = e^{\I \varphi_\eta^{\vec{\mu_1},\vec{\mu_2}}} a^\dagger_{\vec{\mu}_2}a^\dagger_{\vec{\mu_1}}.
\end{align}
To be concrete, we consider the case of unbound anyons, described by clusters with exactly one element.
Here,
\begin{align}
\label{eqnAnyonAlgebra}
 a_p^\dagger a^\dagger_q =& e^{\I \phi_\eta(p-q)} a_q^\dagger a_p^\dagger, \nonumber\\
 a_p a_q^\dagger =& e^{-\I \phi_\eta(p-q)} a_q^\dagger a_p + \delta(p-q),
\end{align}
where the annihilation operator $a_p$ is the hermitian conjugate of $a_p^\dagger$. 
It is striking that the one-dimensional anyonic algebra depends on the relative momentum instead of providing a fixed statistical phase as familiar from two-dimensional anyons \cite{LeinaasMyrheim1977TheoryOfIdenticalParticles} and the different types of one-dimensional anyons mentioned in the introduction \cite{Pasquier1994ALectureOnTheCalogeroSutherlandModels,BrinkHanssonVasiliev1992ExplicitSolutionToTheNBodyCalogeroProblem,FisherGlazman1996TransportInA1DLuttingerLiquid,BatchelorGuanOelkers2006OneDimensionalInteractingAnyonGasLowEnergyPropertiesAndHaldaneExclusionStatistics, Kundu1999ExactSolutionOfDoubleDeltaFunctionBoseGasThroughAnInteractingAnyonGas,TangEggertPelster2015GroundStatePropertiesOfAnyonsInA1DLattice}.
However, in stark contrast to two dimensions, an unperturbed exchange of one-dimensional particles is a priori impossible, as the particles generally scatter. The anyonic character therefore survives only in the scattering properties of the anyons. 
It is plausible that the statistical phase depends on momentum from the point of view of a scattering process.

Given the momentum space operator algebra of \refEq{eqnAnyonAlgebra}, 
we can ask about the algebra of the real space operators  $\Psi^\dagger(x) = \int_{-\infty}^{\infty} dp \ \frac{e^{\I p x}}{\sqrt{2 \pi}} a^\dagger_p
$. 
We obtain
\begin{align}
\label{eqnQuantumBracketsRealSpace}
&\left\{ \Psi(x), \Psi^\dagger(y)\right\} =\delta{(x-y)} \hspace*{-0.4ex}+ \hspace*{-1ex} \intS{0}{\infty} \hspace*{-0.5ex} dz \frac{2e^{-\frac{z}{|\eta|}}}{|\eta|}  \Psi^\dagger(y-z)  \Psi(x-z),
\nonumber
\\
&\left\{ \Psi^\dagger(x), \Psi^\dagger(y) \right\} =  \intS{0}{\infty} dz \ \frac{2{e^{-\frac{z}{|\eta|}}}}{|\eta|}  \Psi^\dagger(y-z)  \Psi^\dagger(x+z),
\end{align}
where $\left\{\dots,\dots\right\}$ is the anticommutator.
Here, $\lim_{\eta \to 0} \int_0^{\infty} dz /|\eta| \, e^{-z/|\eta|} f(z) = f(0)$ yields the bosonic commutation algebra, while the fermionic anticommutation relations for $\eta \to \infty$ are trivially contained.
If we set $x=y$, we obtain a smeared anyonic Pauli principle in real space represented by
\begin{align}
\label{eqnSmearedPauliPrincipleApp}
\left(\Psi^\dagger(x)\right)^2 =  \int_0^{\infty} dz \ \frac{1}{|\eta|} e^{-z/ |\eta|} \Psi^\dagger(y-z)  \Psi^\dagger(x+z).
\end{align}
Finally, 
it is well known that the concept of statistics in one dimension is fuzzy as there exist several ways to transform between different statistics \cite{VonDelftSchoellerBosonizationForBeginnersRefermionizationForExperts, TangEggertPelster2015GroundStatePropertiesOfAnyonsInA1DLattice}.
Likewise here, there is a generalized Jordan-Wigner transformation from the present anyons to the bosons of the Lieb-Liniger model.
Ultimately, this reflects the fact that the Fock space of anyons naturally is isomorphic to the one of the Lieb-Liniger model (if $\eta \neq \pm \infty$).
To this end, consider the bosonic operators $b$ with the algebra $\left[b_k,b_l^\dagger\right]=\delta(k-l)$ and $\left[b_k,b_l^\dagger\right]=0$ with $k,l\in \mathbb{R}$.
For $\eta \neq \pm \infty$, we define the generalized Jordan-Wigner transformation
\begin{align}
\tilde{a}(j) = \lim_{\epsilon \to 0^+} e^{\I \int_{-\infty}^{j-\epsilon} dk \ b_k^\dagger b_k \phi_\eta(k-j)} b(j).
\end{align}
Calculating the algebra of $\tilde{a}$, we find
$\tilde{a}_j \tilde{a}_k = \tilde{a}_k \tilde{a}_j e^{\I \phi_\eta(k-j)}$ 
and
$\tilde{a}_j \tilde{a}^\dagger_k = \tilde{a}^\dagger_k \tilde{a}_j e^{-\I \phi_\eta(j-k)}+ \delta(j-k)$,
which is exactly the anyonic algebra described in \refEq{eqnAnyonAlgebra}. 
As an apparent peculiarity, we have $\tilde{a}^\dagger_k\tilde{a}_k = b^\dagger_k b_k$, which derives the same free Hamiltonian, \refEq{eqnFreeHamiltonian}, using either the bosonic or the anyonic description. 
One would naively expect, that the transformation should generate the interacting Hamiltonian of \refEq{eqnHamiltonian} instead. However, since the theory of Leinaas and Myrheim is intrinsically constrained to the region $\mathcal{R}$, defined at the beginning of the model section, it is no surprise that the transformation obtains \refEq{eqnFreeHamiltonian}. The important information remains encoded in the boundary conditions rather than in the energy.

\section{Systems of finite size}
\label{sctnFiniteSizeSystems}
When anyons are confined to the length $L$, %
one would expect the Dirichlet boundary conditions 
$
\Psi\left(0,x_2, \dots, x_n\right) = \Psi\left(x_1,  \dots,  x_{n-1},L\right) = 0
$
to quantize the allowed momenta, similar to the particle in a box problem.
In fact, the conditions translate to 
\begin{align}
\label{eqnAlphaConstraintBox}
\alpha\left(-k_1, \dots, k_n\right) =& - \alpha\left(\vec{k}\right) ,
\nonumber
\\
\alpha\left(k_1, \dots, -k_n\right) =& - e^{2\I k_n L} \alpha\left(\vec{k}\right).
\end{align}
These constraints of \refEq{eqnAlphaConstraintBox} are only consistent with Eqs.~(\ref{eqnAlphaConstraint}) and (\ref{eqnCuttingCondition}) if the system of transcendental equations
\begin{align}
\label{eqnMomentumQuantizationBox}
 L k_j + \sumS{1\leq (i\neq j) \leq n}{0.15} %
 \left[  \phi_\eta\left({{k}_i-{k}_j}\right) - \phi_\eta\left({{k}_i+{k}_j}\right) \right]/2 =  \pi z_j
\end{align}
is fulfilled for $j$ between $1$ and $n$.
Here, the $z_j$ are positive integers. The momenta that solve \refEq{eqnMomentumQuantizationBox} are discrete and readily numerically obtainable.
In the context of the Lieb-Liniger model, these equations are very similar to the so-called logarithmic Bethe ansatz equations \cite{BatchelorBortzGuanOelkers2005EvidenceForTheSuperTonksGirardeauGas,BatchelorGuanHe2007TheBetheAnsatzFor1DInteractingAnyons}. Note, however, that the Lieb-Liniger Bethe ansatz equations originally describe particles that are confined to a ring, while \refEq{eqnMomentumQuantizationBox} is adjusted to the case of particles in a box. Although no differences in the thermodynamic limit are to be expected, these Bethe ansatz equations should give more reasonable finite size results for confined particles (see  \citeRef{Gaudin1971BoundaryEnergyOfABoseGasIn1D} and \citeRef{HaoZhangLiangChen2006GroundStatePropertiesOf1DUltracoldBoseGasesInAHardWallTrap} for the discussion in the context of the Lieb-Liniger model).

\begin{figure*} 
	\begin{overpic}[height= 12.1 \baselineskip]{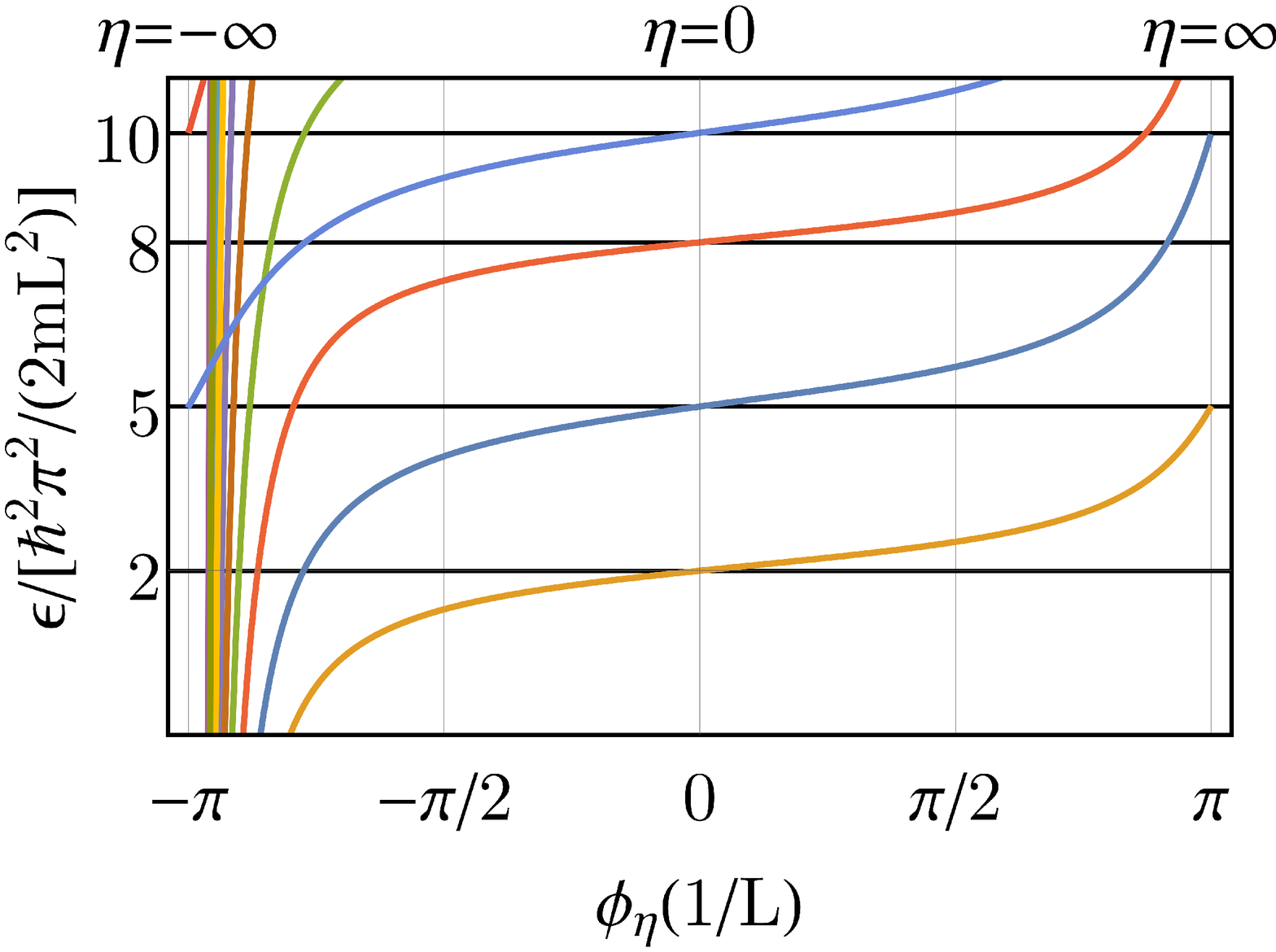}
	\put (-1,66) {\textbf{(a)}}
	\end{overpic}
%
	\begin{overpic}[height= 11.1 \baselineskip]{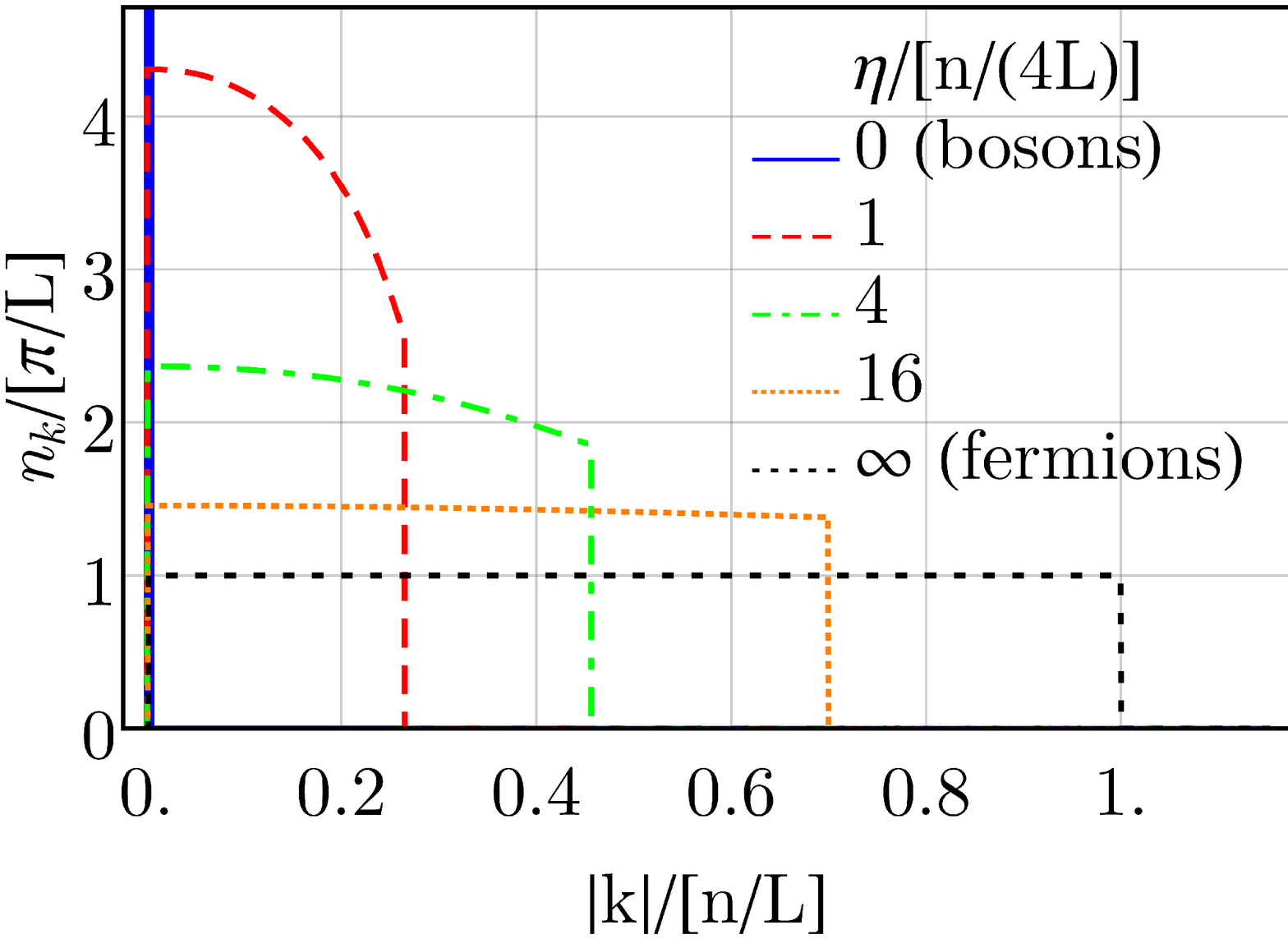}
	\put (-1,67) {\textbf{(b)}}
	\end{overpic}
%
\hspace*{0.5ex}
	\begin{overpic}[height= 11.1 \baselineskip]{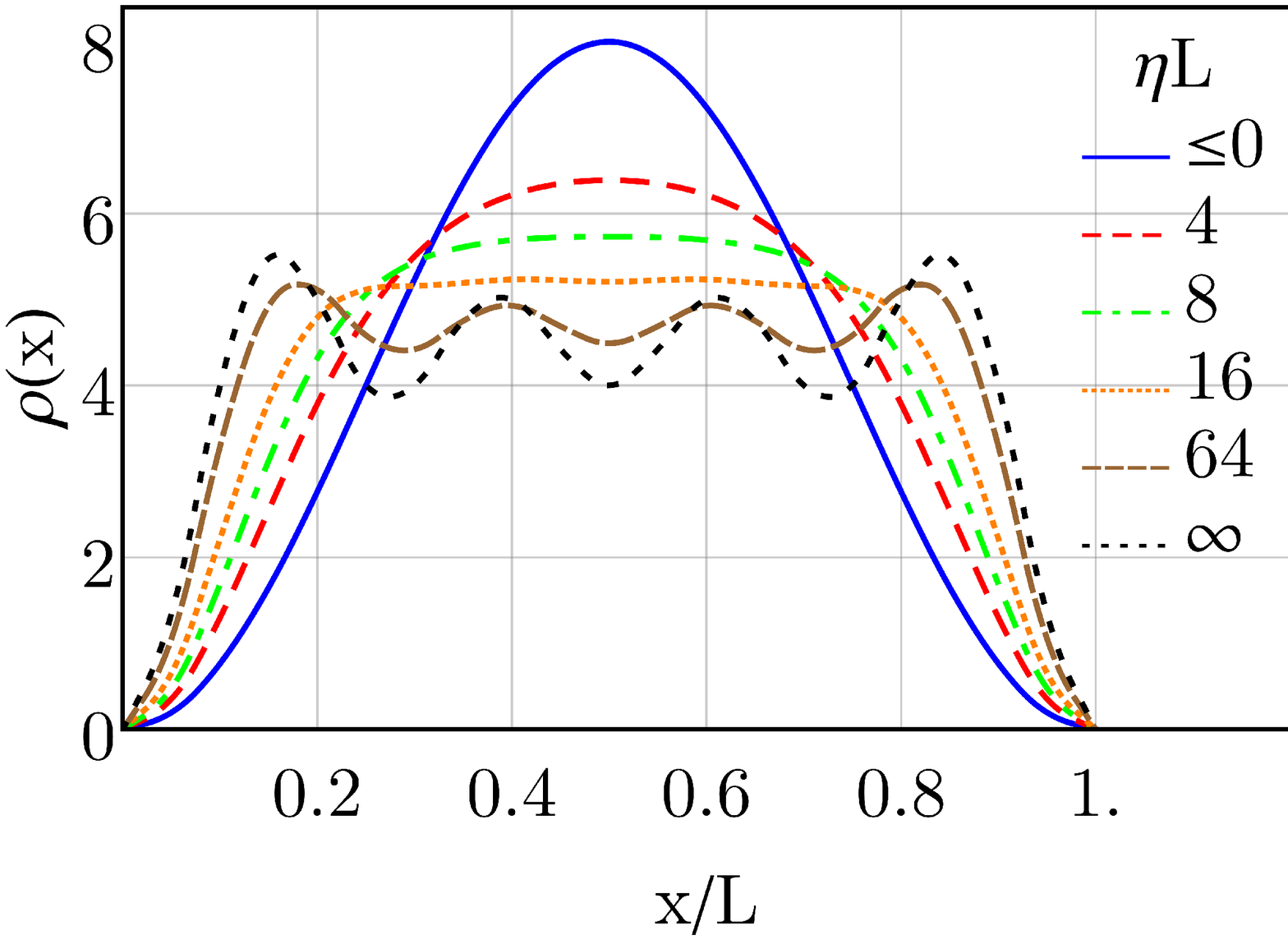}
	\put (-2,67.5) {\textbf{(c)}}
	\end{overpic}
\caption{\label{figResultsAnyonsInABox_all}
Observables for confined anyons. The anyonic properties for $0<\eta<\infty$ continuously interpolate between bosons ($\eta=0$) and fermions ($\eta= \infty$). 
The statistical condensate forms at $\eta<0$.
{(a)} \label{figResultsAnyonsInABox_spectrum} Discrete energy spectrum of two anyons. To depict the full range of $\eta$, we plot against $\phi_\eta(1/ L)$. The two-anyon bound state and its excitations emerge for negative $\eta$. {(b)} \label{figResultsAnyonsInABox_statistics} Momentum density at zero temperature 
in the limit of infinitely many particles (numerical calculations for $n=512$ anyons, where the curves almost converge to the limit $n\to \infty$).
{(c)} \label{figResultsAnyonsInABox_finiteSizeOscis} Finite size oscillations of the particle density $\rho$ for four anyons. 
}
\end{figure*}

\section{Application}
\label{sctnApplication}
Equipped with the developed formalism, we now consider observables of experimental interest.
First, we calculate the spectrum of two confined anyons numerically by solving \refEq{eqnMomentumQuantizationBox}. The result is depicted in \refFig{figResultsAnyonsInABox_all}a.
The anyonic spectra interpolate between the familiar bosonic and fermionic particle-in-a-box spectra for positive $\eta$.
For instance, setting $E_0 = \hbar^2\pi^2/(2mL^2)$, the bosonic level with an energy of $2 E_0$ continuously evolves to the fermionic level with $5 E_0$.
At negative $\eta$, the anyonic levels form two-anyon bound states with an energy proportional to $-\eta^2$ in the infinite-size limit $L\to \infty$.
Energetically higher anyonic bound states correspond to kinetic excitations of the composite anyon in analogy to the behavior of a single particle in a box.
Some anyonic levels refuse to form bound states and instead converge to fermionic energies as $\eta\to -\infty$. These levels ensure that the finite-size spectrum coherently converges to the infinite-size spectrum. %
Energy spectra could be a viable observable in systems with few anyons, like, potentially, interacting cold-atom chains 
\cite{%
	KeilmannLanzmichMcCullockRoncaglia2011StatisticallyInducedPhaseTransitionsAndAnyonsIn1DOpticalLattices,%
	TangEggertPelster2015GroundStatePropertiesOfAnyonsInA1DLattice,%
	HaoZhangChen2009GroundStatePropertiesOfHardCoreAnyonsIn1DOpticalLattices%
}, and detectable by spectroscopic techniques. 
Turning to systems containing many anyons, as possibly the case in solid state systems, 
unavoidable level broadening renders an accurate measurement of the discrete spectrum unfeasible; yet the momentum distribution could uncover the character of the anyons \cite{TangEggertPelster2015GroundStatePropertiesOfAnyonsInA1DLattice}. We depict the momentum density $n_k$ at zero temperature in \refFig{figResultsAnyonsInABox_all}b. This function gives the number of anyons with momentum between $k_1$ and $k_2$ by $\int_{k_1}^{k_2} \hspace*{-0.5 ex} dk\, n_k$.
For bosons and fermions, it is proportional to the Bose-Einstein and Fermi-Dirac distribution, respectively. Anyons with a positive statistical parameter transform these distributions into each other, still remaining a sharply defined chemical potential reflected by a discontinuity in $n_k$. This has to be seen in contrast to the behavior of a Tomonaga-Luttinger liquid.
If the spectral properties of a system are inaccessible, the statistics is still inferable via local properties, e.g., the finite size density fluctuations \cite{KeilmannLanzmichMcCullockRoncaglia2011StatisticallyInducedPhaseTransitionsAndAnyonsIn1DOpticalLattices,StraeterSrivastavaEckardt2016FloquetRealizationAndSignaturesOf1DAnyonsInAnOpticalLattice}. 
While bosons condense to the middle of the system, fermions distribute equally spaced (by Pauli repulsion), resulting in oscillations of the particle density.
Figure \ref{figResultsAnyonsInABox_finiteSizeOscis}c depicts the scenario for four anyons in the ground state.
Unbound anyons suppress the fermionic peaks and broaden the bosonic one, which is characteristic to intermediate statistics \cite{KeilmannLanzmichMcCullockRoncaglia2011StatisticallyInducedPhaseTransitionsAndAnyonsIn1DOpticalLattices,StraeterSrivastavaEckardt2016FloquetRealizationAndSignaturesOf1DAnyonsInAnOpticalLattice}.

\section{The quantum-statistical condensate}
\label{sctnTheQuantumStatisticalCondensate}
For $\eta<0$, the anyonic ground state is a cluster of the form $\mu_j = \I \eta [j - (n-1)/2]$ as depicted in \refFig{figGraphicalRepresentationOfBoundStates} \footnote{This can be seen by considering a general basis function in momentum space $\vec{k} = \left(\vec{\mu}^1,\dots,\vec{\mu}^m \right)$ that consists of $m$ independent clusters and plugging the state into \refEq{eqnHamiltonian}. The energy contribution of clusters separates such that resting clusters and large clusters are generally energetically preferred.}
We call this cluster the quantum-statistical condensate since, in the anyonic picture, its origin is purely based on the quantum statistics. 
It can be conceived as a single composite anyon and corresponds to the bound state of bosons that forms in the Lieb-Liniger model \cite{MontinaArecchi2005ManyBodyGroundStatePropertiesOfAnAttractiveBoseEinsteinCondensateInA1DRing,KanamotoSaitoUeda2003QuantumPhaseTransitionIna1DBoseEinsteinCondensateWithAttractiveInteractions} for an attractive interaction.
Therefore, its local density is similar to the one of as a single quantum particle, which in turn is the same as the one of the Bose-Einstein condensate, cf.  \refFig{figResultsAnyonsInABox_finiteSizeOscis}c.
In fact, the Bose-Einstein condensate can be interpreted as the limit of the quantum-statistical condensate as $\eta \to 0^-$. 
Besides this, both condensates differ profoundly: bosons condense into their {single-particle} ground state, but anyons into an inseparable {many-body} ground state. 
Let us derive further characteristics of the quantum-statistical condensate.
First, we obtain its ground state energy 
\begin{align}
\epsilon_{\text{GS}}= - \frac{\hbar^2}{24 m}\eta^2(n-1)n(n+1)
\end{align}
by \refEq{eqnHamiltonian}.
The proportionality to $n^3$ reveals an exceptional stability of the condensate \footnote{This has first been wrongly described in \citeRef{LiebLiniger1963} but corrected in \citeRef{McGuire1964StudyOfExactlySoluble1DNBodyProblems}.}. 
Let us, for a moment, regard charged anyons exhibiting Hubbard repulsion, which has an associated energy proportional to $n^2$. 
Then, providing a sufficiently large number of anyons, the negative statistical energy outperforms the positive one created by charge repulsion.
The quantum-statistical condensate is hence stable against the introduction of charge.
We conjecture that this property could lead to anyon superconductivity \cite{Laughlin1988SuperconductingGroundStateOfNoninteractingParticlesObeyingFractionalStatistics,Wilczek1990FractionalStatisticisAndAnyonSuperconductivity,BisharaNayak2007NonAbelianAnyonSuperconductivity}. 
In fact, in the context of the Lieb-Liniger model, a phase of pairwisely bound bosons has been predicted \cite{BatchelorBortzGuanOelkers2005EvidenceForTheSuperTonksGirardeauGas}.
A cluster behaves as an individual anyon, the energy of which separates into a kinetic and an internal part. 
Additionally, by \refEq{eqnQuantumAlgebraClusters}, clusters acquire different statistical phases than their constituents.  
For instance, clusters of two anyons behave like anyons with the statistical phase $2\phi_\eta + \phi_{2\eta}$, see \refApp{appBehaviorOfClusters} for details. 
In the vocabulary of topological field theories for two-dimensional anyons  \cite{Kitaev2006AnyonsInAnExactlySolvableModelAndBeyond,TrebstTroyerWangLudwig2008AShortIntroductionToFibonacciAnyonModels,NayakSternFreedmanDasSarma2008NonAbelianAnyonsAndTopologicalQuantumComputation}, the formation of clusters is linked to anyon fusion \label{pageFusionRules}
\footnote{
We relate to the physically motivated concept of anyon fusion. A rigorous mathematical mapping of our theory to a topological field theory is not straightforwardly given.
}.

\section{Conclusions}
\label{sctnConclusions}
On the basis of the general assumptions of Leinaas and Myrheim \cite{LeinaasMyrheim1977TheoryOfIdenticalParticles}, we derive an exact quantum many-body formalism for one-dimensional anyons 
including the exact wave functions, the second quantization, and the momentum discretizing equations for anyons in a box.
The formalism is based on the equivalence to the Lieb-Liniger model of $\delta$-function interacting bosons for which an interpretation in the anyonic context is established.
We numerically calculate characteristic observables, namely, the energy spectrum, the momentum statistics, and the finite-size density fluctuations.
For a negative statistical parameter, anyons attract each other with a force purely induced by their quantum statistics and form the quantum-statistical condensate.
This genuine quantum many-body phase is more robust than the Bose-Einstein condensate. 
In particular, the statistical condensate is stable for charged anyons in the presence of Coulomb repulsion. The clusters themselves should be conceived as individual anyons themselves and obtain a different statistical phase than their constituents.
Our work shows that one-dimensional anyons exhibit original and interesting physics even in the absence of exchangeability.
Furthermore it emphasizes the link between anyons and interacting bosons and thereby opens new possibilities of synthesizing either physical system by its equivalent partner.%
\begin{acknowledgments}
We would like to thank 
Thors Hans Hansson and
Jon Magne Leinaas 
for their clarifying remarks.
Additionally, we acknowledge interesting discussions with
Pablo Burset, 
Fran{\c{c}}ois Cr{\'e}pin, 
Daniel Hetterich,
Axel Pelster, and
Nils Rosehr.
BT thanks the DFG for financial support through the SFB 1170 ("ToCoTronics").
\end{acknowledgments}

\appendix

\section{Notions of intermediate statistics in one spatial dimension}
\label{appDifferentFormalismsForAnyons}
There exists a variety of formalisms describing particles of intermediate statistics in one dimension, which are expected to be applicable to different physical situations.
Albeit they differ in their phenomenology, these particles  are all occasionally called anyons.
For clarity, we shortly discuss the most broadly known theories that are applicable to one spatial dimension.

If the occupation number of a single particle quantum state is restricted to maximally assume a given integer, the particles can be described as parafermions, which are closely related to Potts and clock models \cite{Fendley2012ParafermionicEdgeZeroModesInZNInvariantSpinChains,AliceaFendley2016TopologicalPhasesWithParafermionsTheoryAndBlueprints} and Gentile statistics \cite{Gentile1940OsservazioniSopraLeStatisticheIntermedie}.
Such particles are, amongst others, expected to exist as magnetic excitations \cite{DaiXie2009IntermediateStatisticsSpinWaves,HutterWoottonLoss2015ParafermionsInAKagomeLatticeOfQuibitsForTopologicalQuantumComputation}.
Another kind of intermediate statistics considers the representations of the local current algebra (the commutation relations between the particle density and the particle currents in all spatial dimensions) \cite{%
	GoldinMenikoffSharp1980ParticleStatisticsFromInducedRepresentationsOfALocalCurrentGroup,%
	GoldinMenikoffSharp1981RepresentationsOfALocalCurrentAlgebraInSonsimplyConnectedSpaceAndTheAharonovBohmEffect%
} 
or the quantization of the algebra of allowed observables of indistinguishable particles. The latter has been applied to superconducting vortices \cite{LeinaasMyrheim1988IntermediateStatisticsForVorticesInSuperfluidFilms} and two-dimensional anyons effectively confined to one dimension by a strong magnetic field \cite{Hansson1992DimensionalReductionInAnyonSystems}.
Yet another notion of anyons in one dimension can be derived from Haldane's generalization of the Pauli principle \cite{Haldane1991FractionalStatisticsInArbitraryDimensionsAGeneralizationOfThePauliPrinciple}, which is, for instance, applicable to spinon excitations in spin chains. In this approach, the single-particle Hilbert space dimension depends on the total number of particles in the system.
Finally, the name anyons is used in one dimension to describe low-energy quasi-particle excitations of interacting fermionic systems \cite{Ha1995FractionalStatisticsInOneDimensionViewFromAnExactlySolvableModel} linked to the Calogero-Sutherland model \cite{Pasquier1994ALectureOnTheCalogeroSutherlandModels,BrinkHanssonVasiliev1992ExplicitSolutionToTheNBodyCalogeroProblem}, the Haldane-Shastry chain, and the fractional excitations in Tomonaga-Luttinger liquids \cite{FisherGlazman1996TransportInA1DLuttingerLiquid,BatchelorGuanOelkers2006OneDimensionalInteractingAnyonGasLowEnergyPropertiesAndHaldaneExclusionStatistics, Kundu1999ExactSolutionOfDoubleDeltaFunctionBoseGasThroughAnInteractingAnyonGas,TangEggertPelster2015GroundStatePropertiesOfAnyonsInA1DLattice}.
It is known that these particles (considering each channel separately in the case of a Tomonaga-Luttinger liquid) break time reversal symmetry on the fundamental level of their operator algebra, 
reflected by an asymmetric momentum distribution \cite{HaoZhangChen2009GroundStatePropertiesOfHardCoreAnyonsIn1DOpticalLattices,TangEggertPelster2015GroundStatePropertiesOfAnyonsInA1DLattice}.

\section{Interpretation of anyon clusters as individual anyons}
\label{appBehaviorOfClusters}
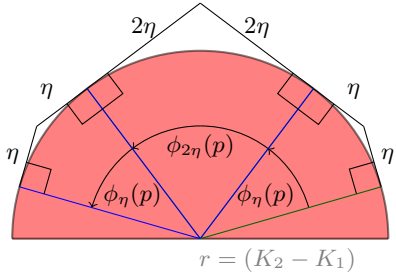
\begin{figure}

\definecolor{darkgreen}{rgb}{0,0.4,0}
{
\begin{tikzpicture}[scale = 2.5]
\clip	(-1.1,-0.2)	rectangle	(1.1,1.35);
\filldraw[thick,fill=red, opacity=0.5]	node[anchor=north west,]{\hspace*{-1ex}$r=\left(K_2 - K_1\right)$} (1,0)	arc 	(0:180:1) ;
\draw[thick,fill=red, opacity=0.5]	(-1,0) 	--	(1,0)	;

\def\ETA{37}
\def\angleSize{0.6};

\begin{scope}[rotate= \ETA*3/2]
	\def\t{\ETA/2};
	\draw[->]	(0,\angleSize)	arc 	(90:{90+\t}:\angleSize);
	\draw	(0,\angleSize)	arc 	(90:{90-\t}:\angleSize);
	\node[anchor= north west] at (0,\angleSize) {\hspace*{-2ex} $\phi_\eta(p)$};
	\path [name path=radius1] 	(0,0)	--	(-2,{2*cot(\t)});
	\path [name path=circle1]	(0,0)	circle	(1);
	\draw [color=blue, name intersections={of=radius1 and circle1, by={s1}}]	(0,0)	--	(s1); 
	\path [name path=radius2] 	(0,0)	--	(2,{2*cot(\t)});
	\draw [color=darkgreen, name intersections={of=radius2 and circle1, by={s2}}]	(0,0)	--	(s2); 
	\path [name path=momentum]	(0,0)	--	(0,{tan(\t)+1});
	\path [name path=tangent1]	(s1)	--	($(1,{tan(\t)} )+(s1)$);
	\draw [name intersections={of=momentum and tangent1, by={eta1}}] (s1)	--	node[anchor=east]{$\eta$}	(eta1);
	\draw (s2)	--	node[anchor=south east]{$\eta$}	(eta1);
	\draw [right angle symbol={s1}{eta1}{0,0}];
	\draw [right angle symbol={s2}{eta1}{0,0}];
\end{scope}

\begin{scope}
	\def\t{\ETA};
	\draw[->]	(0,\angleSize)	arc 	(90:{90+\t}:\angleSize);
	\draw	(0,\angleSize)	arc 	(90:{90-\t}:\angleSize);
	\node[anchor=north] at (0,\angleSize) {$\phi_{2\eta}(p)$};
	\path [name path=radius1] 	(0,0)	--	(-2,{2*cot(\t)});
	\path [name path=circle1]	(0,0)	circle	(1);
	\draw [color=blue, name intersections={of=radius1 and circle1, by={s1}}]	(0,0)	--	(s1); 
	\path [name path=radius2] 	(0,0)	--	(2,{2*cot(\t)});
	\draw [color=darkgreen, name intersections={of=radius2 and circle1, by={s2}}]	(0,0)	--	(s2); 
	\path [name path=momentum]	(0,0)	--	(0,{tan(\t)+1});
	\path [name path=tangent1]	(s1)	--	($(1,{tan(\t)} )+(s1)$);
	\draw [name intersections={of=momentum and tangent1, by={eta1}}] (s1)	--	node[anchor=south]{$2 \eta$}	(eta1);
	\draw (s2)	--	node[anchor=south]{$2 \eta$}	(eta1);
	\draw [right angle symbol={s1}{eta1}{0,0}];
	\draw [right angle symbol={s2}{eta1}{0,0}];
\end{scope}

\begin{scope}[rotate=-\ETA*3/2]
	\def\t{\ETA/2};
	\draw[->]	(0,\angleSize)	arc 	(90:{90+\t}:\angleSize);
	\draw	(0,\angleSize)	arc 	(90:{90-\t}:\angleSize);
	\node[anchor=north east] at (0,\angleSize) {$\phi_\eta(p)$\hspace*{-1ex}};
	\path [name path=radius1] 	(0,0)	--	(-2,{2*cot(\t)});
	\path [name path=circle1]	(0,0)	circle	(1);
	\draw [color=blue, name intersections={of=radius1 and circle1, by={s1}}]	(0,0)	--	(s1); 
	\path [name path=radius2] 	(0,0)	--	(2,{2*cot(\t)});
	\draw [color=darkgreen, name intersections={of=radius2 and circle1, by={s2}}]	(0,0)	--	(s2); 
	\path [name path=momentum]	(0,0)	--	(0,{tan(\t)+1});
	\path [name path=tangent1]	(s1)	--	($(1,{tan(\t)} )+(s1)$);
	\draw [name intersections={of=momentum and tangent1, by={eta1}}] (s1)	--	node[anchor=south west]{$\eta$}	(eta1);
	\draw (s2)	--	node[anchor=west]{$\eta$}	(eta1);
	\draw [right angle symbol={s1}{eta1}{0,0}];
	\draw [right angle symbol={s2}{eta1}{0,0}];
\end{scope}
\end{tikzpicture}
}
\caption{\label{figGeometricalPhaseShiftTwoBoundStates}Geometric interpretation of the statistical phase of two clusters, each consisting of two anyons. The radius of the circle denotes the relative momentum between clusters $K_2 - K_1$. The statistical angle is obtained by adding up three summands. Two of these summands are the normal statistical angle $\phi_\eta$, the third summand is the statistical angle of the doubled statistical parameter $\phi_{2\eta}$. The statistical parameter $\eta$ appears in the lengths of the drawn tangential segments.}
\end{figure}

We want to show how the exchange phase of clusters can be interpreted as the statistical phase of a composite species of anyons reaching further than the interpretation supported by \refEq{eqnQuantumAlgebraClusters}. 
To this end, we consider two clusters of anyons $\vec{\mu_1} = \left(K_1 + \I \eta/2, K_1 - \I \eta/2\right)$ and $\vec{\mu_2} = \left(K_2 + \I \eta/2, K_2 - \I \eta/2\right)$, the cluster structures of which are depicted in \refFig{figGraphicalRepresentationOfBoundStates}b.
We introduce the center of mass coordinates $X_1 = \left(x_1 + x_2\right) /2$ and $X_2 = \left(x_3 + x_4 \right)/2$, as well as the relative coordinates $Z_1 = \left(x_2 - x_1 \right)/2$ and $Z_2 = \left(x_4 -x_3\right)/2$.
Under the assumption that the two clusters are sufficiently far away from each other, i.e., $X_2 - X_1 \to \infty$ and $Z_1,Z_2$ finite, we obtain 
\begin{align}
&\Psi_{\left(\mu_1,\mu_2\right)}(X_1,X_2,Z_1,Z_2) \propto  
\nonumber 
\\ 
	&\left[
		e^{2 \I \left( K_1 X_1 + K_2 X_2\right)} +
		e^{\I \varphi_\eta^{\vec{\mu_1},\vec{\mu_2}}} e^{2 \I \left( K_2 X_1 +  K_1 X_2\right)}
	\right]
e^{\eta \left(Z_1 + Z_2 \right)}.
\end{align}
This wave function obtains the form of a wave function of two composite anyons with an altered statistical phase of $\varphi_\eta^{\vec{\mu_1},\vec{\mu_2}}$, especially if we recall that $Z_1$ and $Z_2$ are of the order of $1/\eta$. 
This can be physically interpreted as the fusion of anyons to clusters which themselves behave as a composite anyon species.
Interestingly, the new statistical phase is
\begin{align}
\varphi_\eta^{\vec{\mu_1},\vec{\mu_2}}
= 2 \phi_\eta(K_2-K_1) + \phi_{2\eta}({K_2-K_1}),
\end{align}
where $\phi_\eta$ is the statistical phase defined in \refEq{eqnStatisticalPhase}.
This has an appealing geometric interpretation, 
which we depict in \refFig{figGeometricalPhaseShiftTwoBoundStates}.

%

\end{document}